\documentclass[aps,prl,twocolumn,superscriptaddress,longbibliography]{revtex4-2}
\usepackage{breakurl}
\usepackage{xcolor, soul}
\usepackage{sidecap}
\usepackage{amssymb}
\usepackage{hhline}
\usepackage{multirow}
\sidecaptionvpos{figure}{t}
\usepackage{amsmath}
\usepackage{graphicx}
\usepackage{esint}
\usepackage{epstopdf}
\usepackage{rotating}
\usepackage{framed}
\graphicspath{{pict/}{./}}
\usepackage{bm}%
\usepackage{microtype,bm,bbm,graphicx,booktabs,times}

\newcounter{Fig}

\newcommand\mymapstol{\mathrel{\ooalign{$\leftarrow$\cr%
  \kern1.75ex\raise0.275ex\hbox{\scalebox{1}[0.4]{$\mid$}}\cr}}}

\newcommand\mymapstor{\mathrel{\ooalign{$\rightarrow$\cr%
  \kern-.15ex\raise.275ex\hbox{\scalebox{1}[0.4]{$\mid$}}\cr}}}
  
\usepackage{graphicx,tipa}

\usepackage{tikz}

\makeatletter
\newcommand*{\rom}[1]{\expandafter\@slowromancap\romannumeral #1@}
\makeatother

\begin{document}

\title{Quasi-normal modes empowered coherent control of electromagnetic interactions}
\author{Jingwei Wang}
\affiliation{School of Optical and Electronic Information, Huazhong University of Science and Technology, Wuhan, Hubei 430074, P. R. China}
\author{Pengxiang Wang}
\affiliation{School of Optical and Electronic Information, Huazhong University of Science and Technology, Wuhan, Hubei 430074, P. R. China}
\author{Chaofan Zhang}
\email{c.zhang@nudt.edu.cn}
\affiliation{College for Advanced Interdisciplinary Studies, National University of Defense Technology, Changsha 410073, P. R. China.}
\affiliation{Nanhu Laser Laboratory and Hunan Provincial Key Laboratory of Novel Nano-Optoelectronic Information Materials and
Devices, National University of Defense Technology, Changsha 410073, P. R. China.}
\author{Yuntian Chen}
\email{yuntian@hust.edu.cn}
\affiliation{School of Optical and Electronic Information, Huazhong University of Science and Technology, Wuhan, Hubei 430074, P. R. China}
\affiliation{Wuhan National Laboratory for Optoelectronics, Huazhong University of Science and Technology, Wuhan, Hubei 430074, P. R. China}
\author{Wei Liu}
\email{wei.liu.pku@gmail.com}
\affiliation{College for Advanced Interdisciplinary Studies, National University of Defense Technology, Changsha 410073, P. R. China.}
\affiliation{Nanhu Laser Laboratory and Hunan Provincial Key Laboratory of Novel Nano-Optoelectronic Information Materials and
Devices, National University of Defense Technology, Changsha 410073, P. R. China.}

\begin{abstract}
Quasi-normal modes (QNMs) and coherent control of light-matter interactions (through synchronized multiple coherent incident waves) are profound and pervasive concepts in and beyond photonics, making accessible photonic manipulations with extreme precision and efficiency. Though each has been playing essential roles in its own, these two sweeping concepts remain largely segregated with little interactions, blocking vast opportunities of cross fertilization to explore. Here we unify both concepts into a novel framework of coherent control for light interacting with open photonic systems. From the QNM perspective, scattered waves are superimposed radiations from all QNMs excited, and thus coherent controls can be mapped into another problem of QNM excitation manipulations. Within our framework, all incident properties (amplitudes, phases and polarizations) of waves from different directions can be exploited simultaneously in a synchronous manner, facilitating independent manipulations of each QNM and thus unlocking enormous flexibilities for coherent controls of both scattering intensity and polarization: (i) A visible structure under a single incident wave can be made invisible through shining extra waves; (ii) Along a direction where QNMs' radiation polarizations are identical, scattering along this direction can be fully eliminated, thus generalizing Kerker effects from a distinct QNM perspective; (iii) Along a direction of distinct QNM radiation polarizations, arbitrary scattering polarizations can be obtained.  Given the ubiquity and profundity of QNMs and coherent control in almost all branches of wave physics, our framework and its underlying principles will inspire further fundamental explorations and practical applications beyond photonics, opening new opportunities for various forms of wave-matter interactions. 
\end{abstract}

\maketitle

Efficient manipulations of light-matter interactions constitute the cornerstone of many branches of physical sciences, which up to now largely rely on either structuring matter (both chemically and geometrically, such as in \textit{e.g.} the field of metamaterials and its associated disciplines~\cite{PENDRY_2006_Science_Controlling,Cai2010_book,YU_NatMater_flat_2014,LIU_2018_Opt.Express_Generalized,BABICHEVA_2024_Adv.Opt.Photon.AOP_Mieresonanta}) and/or structuring light~\cite{TANG_Phys.Rev.Lett._Optical,FORBES_2021_Nat.Photonics_Structured,BLIOKH_2023_J.Opt._Roadmap,SHEN_2024_Nat.Photon._Optical}. For the latter category, coherent control of light-matter interactions forms a major theme, which employs multiple incident waves from different directions to achieve various exotic functionalities relying on their coherent interactions with photonic structures~\cite{BARANOV2017Nat.Rev.Mater.,wan_timereversed_2011,zhang_controlling_2012}, including finite particles~\cite{XI_2017_Phys.Rev.Lett._Magnetic,BAG_2018_Phys.Rev.Lett._Transverse}, lattices~\cite{ZHANG_2020_AdvancedMaterials_Coherent,KANG_2022_NatCommun_Coherent}, and other random/complex media~\cite{DELHOUGNE_2021_Phys.Rev.Lett._Coherent,JIANG_2024_Nat.Phys._Coherent}, covering not only linear and nonlinear~\cite{RODRIGO_2013_Phys.Rev.Lett._Coherent,SUWUNNARAT_2022_CommunPhys_Nonlinear}, classical and quantum regimes~\cite{JEFFERS_2019_Phys.Rev.Lett._Nonlocal}, and waves of other forms~\cite{MULLERS_2018_Sci.Adv._Coherent}.  Furthermore, topics associated with coherent control are rapidly merging with other vibrant fields such as bound states in the continuum~\cite{KANG_2022_NatCommun_Coherent,WANG_2024_PhotonicsInsights_Optical}, optical singularities~\cite{NYE_natural_1999,ZHUANG_2024_Phys.Rev.Lett._Topological}, generalized Kerker effects~\cite{KUZNETSOV_Science_optically_2016,LIU_2018_Opt.Express_Generalized,BAG_2018_Phys.Rev.Lett._Transverse}, non-Hermitian photonics~\cite{BERRY_2004_CzechoslovakJournalofPhysics_Physicsa,WANG_2023_Adv.Opt.Photon._NonHermitian,WANG_2021_Science_Coherent,HORNER_2024_Phys.Rev.Lett._Coherent}, unlocking many extra degrees of freedom to exploit for more flexible and precise light-matter interaction manipulations.

\begin{figure}[tp]
\centerline{\includegraphics[width=9cm]{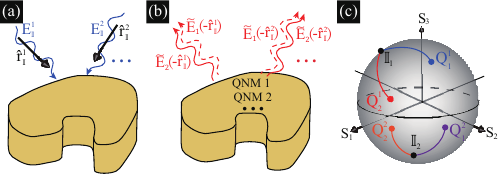}} \caption{\small  (a) A reciprocal photonic scatterer with multiple incident plane waves indexed by $n$: for the $n^ {\mathrm{th}}$ incident wave, the propagation direction unit-vector is $\hat{\mathbf{r}}_{\mathbb{I}}^{n}$ and field vector is $\mathbf{E}_{\mathbb{I}}^n(\hat{\mathbf{r}}_{\mathbb{I}}^{n})$. (b) The scatterer supports a set of discrete QNMs indexed by $m$: the $m^{\mathrm{th}}$ QNM's radiated field along $\mathbf{\hat{r}}$ is $\tilde{\mathbf{E}}_m(\mathbf{\hat{r}})$. The QNM excitation coefficient is only related to QNM radiations opposite to the incident directions [$\tilde{\mathbf{E}}_m(-\hat{\mathbf{r}}_{\mathbb{I}}^{n})$] and has nothing to do with radiations along other directions. (c) On the unit Poincaré sphere, $\mathrm{Q}_m^n$ denotes polarization of the $m^{\mathrm{th}}$ QNM radiations opposite to the $n^ {\mathrm{th}}$ incident direction and  $\mathbb{I}_n$ denotes polarization of the $n^ {\mathrm{th}}$ incident wave. $\mathrm{Q}$ and $\mathbb{I}$ are connected by geodesic great arcs.}
\label{fig1}
\end{figure}

In parallel, another sweeping concept of QNM from a different  field~\cite{KONOPLYA_2011_Rev.Mod.Phys._Quasinormal} pervades rapidly into not only various branches of photonics~\cite{CHING_1998_Rev.Mod.Phys._Quasinormalmode,LALANNE__LaserPhotonicsRev._Light,KRISTENSEN_2020_Adv.Opt.Photon.AOP_Modeling} (fusing with other pervasive concepts such as geometric phase~\cite{WANG_2024_Phys.Rev.Lett._Geometric}, parity-time symmetry~\cite{REN_2021_Phys.Rev.X_Quasinormal}, nonlocality~\cite{ZHOU_2021_Phys.Rev.Lett._General,TAO_2022_Phys.Rev.Lett._Quasinormal}, and polarization singularities~\cite{CHEN_2019__Singularities,CHEN_Phys.Rev.Lett._Extremize}), but also many other disciplines beyond photonics~\cite{JARAMILLO_2021_Phys.Rev.X_Pseudospectrum,JARAMILLO_2022_Phys.Rev.Lett._Gravitational,WILKS_2024_Proc.R.Soc.Math.Phys.Eng.Sci._Generalized}. Nevertheless, up to now the two flourishing fields of coherent control and QNM have been developing almost independently along their own veins with little if not no cross interactions, and thus it is unclear what extra hidden degrees of freedom  and opportunities can be opened if they are merged into a single unified framework. 

Aiming to bridge both concepts, we revisit the seminal subject of coherent control of light-matter interactions, now from the novel perspective of tailored excitations of QNMs with multiple incident plane waves from different directions. For reciprocal open photonic structures, the complex excitation coefficient for each QNM can be directly calculated in the far field in a simple and elegant manner without involving cumbersome near-field integrations~\cite{WANG_2024_Phys.Rev.Lett._Geometric,CHEN_Phys.Rev.Lett._Extremize}. Since scattered waves are coherent superpositions of radiations by all QNMs excited, direct far-field manipulations of QNM excitations provide full flexibilities for coherent control of various interaction properties. We have shown that an originally visible structure with a single incident wave can be made fully invisible when extra tailored waves are shone. And when multiple QNMs are co-excited, for a direction of identical (distinct) QNM radiation polarizations,  scattering along this direction can be made null (of arbitrary polarizations). All those manipulations are achieved through synchronizing the polarizations, phases and amplitudes of the incident plane waves according to a pre-designed recipe provided by our model. The established framework fusing the two profound concepts can empower fundamental explorations and practical applications not only in photonics, but also in many other disciplines that involve matter-wave interactions of various forms.

\section*{Results}
\subsection*{Theoretical Model of coherent QNM excitations}
The optical responses of open photonic structures can be calculated through the discrete set of QNMs they support (indexed by integer $m$), which are characterized by complex eigenfrequencies $\tilde{\omega}$ and eigenfields $\tilde{\mathbf{E}}_m(\mathbf{r})$~\cite{LALANNE__LaserPhotonicsRev._Light,KRISTENSEN_2020_Adv.Opt.Photon.AOP_Modeling}. In the far field, the radiations [$\tilde{\mathbf{E}}_m(\mathbf{\hat{r}})$ with $\mathbf{\hat{r}}$ as the unit direction vector] of each QNM are transverse [$\tilde{\mathbf{E}}_m(\mathbf{\hat{r}})\cdot \mathbf{\hat{r}}=0$] and the radiation polarization can be described by either the Jones vector or Stokes vector $\mathbf{S}$ (with three components $S_{1,2,3}$) on the Poincaré sphere~\cite{YARIV_2006__Photonics}. For coherent control with multiple incident waves (indexed by integer $n$), in the far field the transverse scattered waves $\mathbf{E}_{\rm{sca}}(\hat{\mathbf{r}})$ can be obtained through  coherent additions of radiations from all QNMs excited:
\begin{equation}
\label{expansion}
\mathbf{E}_{\rm{sca}}(\hat{\mathbf{r}})=\sum_m \sum_n \alpha_{m}^{n} {\tilde{\mathbf{E}}}_{m}(\hat{\mathbf{{r}}})=\sum_m \alpha_{m}{\tilde{\mathbf{E}}}_{m}(\hat{\mathbf{{r}}}),  
\end{equation}
where $\alpha_{m}^n$ is the  complex excitation (coupling) coefficient for the $m^ {\mathrm{th}}$ QNM under the $n^ {\mathrm{th}}$ incident wave (the incident unit direction vector is $\hat{\mathbf{r}}_{\mathbb{I}}^{n}$; the subscript $\mathbb{I}$ denotes incidence, as is the case throughout the paper).   For the simple scenario of reciprocal photonic structures and incident plane waves [field vector $\mathbf{E}_{\mathbb{I}}^n(\hat{\mathbf{r}}_{\mathbb{I}}^{n})$;  see Fig.~\ref{fig1}(a)],  the excitation coefficients can be expressed as~\cite{WANG_2024_Phys.Rev.Lett._Geometric}:
\begin{equation}
\label{coefficient-geometric}
\alpha_{m}^{n}\propto \tau_{m}^{n}\cos\left(\frac{1}{2}\overset{\smash{\raisebox{0.01ex}{\tikz\draw (0,0) arc[start angle=120,end angle=60,radius=0.7cm];}}}{\mathbb{I}_n \mathrm{Q}_m^n}\right)\exp({i\varphi_m^n})= \tau_{m}^{n}\cos\left(\frac{1}{2}g_m^n\right)\exp({i\varphi_m^n}).
\end{equation}
Here $\tau_{m}^{n}=|\mathbf{E}_{\mathbb{I}}^n(\hat{\mathbf{r}}_{\mathbb{I}}^{n})|\cdot|\tilde{\mathbf{E}}_m(-\hat{\mathbf{r}}_{\mathbb{I}}^{n})|$ is the amplitude product of the incident field and the QNM radiation field opposite to the incident direction  [Fig.~\ref{fig1}(b)]; $\mathbb{I}_n$ and $\mathrm{Q}_m^n$ are points on the unit Poincaré sphere, denoting respectively the incident polarization for $n^{\mathrm{th}}$ incident wave and the radiation polarization along the opposite incident direction ($\hat{\mathbf{r}}=-\hat{\mathbf{r}}_{\mathbb{I}}^{n}$) from the $m^ {\mathrm{th}}$ QNM  [Fig.~\ref{fig1}(b)]; $g_m^n=\overset{\smash{\raisebox{-0.1ex}{\tikz\draw (0,0) arc[start angle=110,end angle=70,radius=0.9cm];}}}{\mathbb{I}_n \mathrm{Q}_m^n}$ is the geodesic distance between the two points  [Fig.~\ref{fig1}(c)]; $\exp({i\varphi_m^n})$ is the overall phase factor, which is related not only to the phase of the incident wave, but also to the original phase assigned to each QNM~\cite{WANG_2024_Phys.Rev.Lett._Geometric}. We note that the term $g_m^n$ characterizes the polarization difference between the incident wave and the QNM radiation opposite to the incident direction: $g_m^n=0$ ( $\mathbb{I}_n$ and $\mathrm{Q}_m^n$ are overlapped) and $g_m^n=\pi$ ($\mathbb{I}_n$ and $\mathrm{Q}_m^n$ are diametrically opposite antipodal points) represent fully matched and orthogonal polarizations, respectively. It is clear that all the incident information (amplitudes, phases and polarizations) are embedded into the above equation, and the excitation coefficients have nothing to do with QNM radiations along directions not opposite to the incident direction, which not only greatly simplifies the calculations but also more importantly reveals intuitive and elegant principles. 

The physical meaning of Eq.~(\ref{coefficient-geometric}) is consequently quite clear: the amplitude term of the excitation coefficient $\tau_{m}^{n}\cos(\frac{1}{2}\overset{\smash{\raisebox{-0.1ex}{\tikz\draw (0,0) arc[start angle=110,end angle=70,radius=0.9cm];}}}{\mathbb{I}_n \mathrm{Q}_m^n})$ tells that to maximally excite the $m^ {\mathrm{th}}$ QNM with the $n^ {\mathrm{th}}$ incident wave, we should shine the wave along the direction opposite to which the QNM radiation [$\tilde{\mathbf{E}}_m(-\hat{\mathbf{r}}_{\mathbb{I}}^{n})$] reaches its maximum with matched polarization [$g_m^n=0$ and $\cos\left(\frac{1}{2}g_m^n\right)=1$], and the QNM would not be excited when  $\tilde{\mathbf{E}}_m(-\hat{\mathbf{r}}_{\mathbb{I}}^{n})=0$ or the incident polarization is orthogonal to the QNM radiation polarization  [$g_m^n=\pi$ and $\cos\left(\frac{1}{2}g_m^n\right)=0$]~\cite{WEN_2024_Laser&PhotonicsReviews_Momentum}; the excitation phase ($\varphi_m^n$) can be controlled through tuning the incident phase [$\varphi_m^n\rightarrow\varphi_m^n+\varphi_0$ with $\mathbf{E}_{\mathbb{I}}^n(\hat{\mathbf{r}}_{\mathbb{I}}^{n})\rightarrow\mathbf{E}_{\mathbb{I}}^n(\hat{\mathbf{r}}_{\mathbb{I}}^{n})\exp{(i\varphi_0)}$]. Those  tell that the coherent control of light-matter interactions can be  flexibly implemented by tailoring the incident properties solely, through incident amplitudes and polarizations that tune $\tau_{m}^{n}\cos(\frac{1}{2}\overset{\smash{\raisebox{-0.1ex}{\tikz\draw (0,0) arc[start angle=110,end angle=70,radius=0.9cm];}}}{\mathbb{I}_n \mathrm{Q}_m^n})$ or through incident phase that tunes $\varphi_m^n$. We emphasize that for a specific mode radiation polarization  $\mathrm{Q}_m^n$,  a constant $g_m^n=\overset{\smash{\raisebox{-0.1ex}{\tikz\draw (0,0) arc[start angle=110,end angle=70,radius=0.8cm];}}}{\mathbb{I}_n \mathrm{Q}_m^n}=g_0$ ($g_0\neq0,1$) corresponds to all polarizations on a geodesic circle centered around  $\mathrm{Q}_m^n$ with radius $g_0$ on the Poincaré sphere. As a result, any dependence on  $\cos(\frac{1}{2}g)$ corresponds to a correlation with a series of polarizations rather than only one polarization. This is easy to understand~\cite{RAMACHANDRAN_1961}: the incident polarization ${\mathbb{I}}^{n}$ can always be expanded into  $\mathrm{Q}_m^n$-component and orthogonal-$\mathrm{Q}_m^n$-component; $g_m^n$ is only relevant to the relative amplitude of those two components and irrelevant to their relative phase; different relative phases correspond to different points on the geodesic circle around $\mathrm{Q}_m^n$ (refer also to more explicit discussions after Eq.~(\ref{expansion-directional})).

\subsection*{Coherent excitation of an electric dipole}
\begin{figure}[tp]
\centerline{\includegraphics[width=8.5cm]{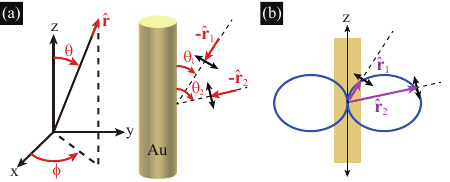}} \caption{\small (a) A metal cylinder supports an ED resonance which is excited by two linearly-polarized plane waves  (the polarization is parallel to the shared incident plane)  along $-\mathbf{\hat{r}}_{1,2}$ that make angles $\theta_{1,2}$  with respect to the cylinder axis. The alongside Cartesian coordinate system is parameterized by azimuthal angle $\phi$ and polar angle $\theta$. (b) The on-plane angular radiation pattern of the ED, and the radiations are also on-plane linearly polarized: along $\mathbf{\hat{r}}_{1,2}$ the radiation polarizations are fully matched to those of the incident waves: $\cos\left(\frac{1}{2}g_{\rm{ED}}^{1,2}\right)=1$.}
\label{fig2}  
\end{figure}

The validity and power of our framework for coherent control can be exemplified with the simplest scenario of excitation of an electric dipole (ED)  supported by metal cylinder with two incident plane waves (\textit{e.g.} linearly polarized on the shared incident plane) from two different directions [opposite to $\hat{\mathbf{r}}_{1,2}$ that makes angles $\theta_{1,2}$ with respect to the \textbf{z} axis; see Fig.~\ref{fig2}(a)]. From a near-field perspective, the ED originates from the harmonic current oscillation along the cylinder axis (\textbf{z} axis), which interacts with electric field component $\mathbf{E}_z$ only. As a result, the ED excitation coefficient $\alpha_{\rm{ED}}$ can be expressed as:
\begin{equation}
\label{coefficient-ED}
\alpha_{\rm{ED}}\propto|\mathbf{E}_{\mathbb{I}}^1(-\hat{\mathbf{r}}_{1})|\cdot|\cos(\theta_1)|+|\mathbf{E}_{\mathbb{I}}^2(-\hat{\mathbf{r}}_{2})|\cdot|\cos(\theta_2)|\Delta\varphi,
\end{equation}
where $\Delta\varphi$ is the phase contrast between the two contributing channels. While through our far-field model, alternatively, according to Eq.~(\ref{coefficient-geometric}) we have:
\begin{equation}
\label{coefficient-ED-far}
\alpha_{\rm{ED}}\propto|\mathbf{E}_{\mathbb{I}}^1(-\hat{\mathbf{r}}_{1})|\cdot|\tilde{\mathbf{E}}_{\rm{ED}}(\hat{\mathbf{r}}_{1})|+|\mathbf{E}_{\mathbb{I}}^2(-\hat{\mathbf{r}}_{2})|\cdot|\tilde{\mathbf{E}}_{\rm{ED}}(\hat{\mathbf{r}}_{2})|\Delta\varphi.
\end{equation}
Since the radiations of ED are also linearly polarized on the radiation plane [see Fig.~\ref{fig2}(b)], we have for both incident waves $g_{\rm{ED}}^{1,2}=0$ and $\cos\left(\frac{1}{2}g_{\rm{ED}}^{1,2}\right)=1$,  through which combined with Eq.~(\ref{coefficient-geometric}) we obtain Eq.~(\ref{coefficient-ED-far}). The consistence between Eqs.~(\ref{coefficient-ED}) and (\ref{coefficient-ED-far}) requires that the angular radiation intensity distribution must observe the relation $\tilde{\mathbf{I}}_{\rm{ED}}(\hat{\mathbf{r}})=|\tilde{\mathbf{E}}_{\rm{ED}}(\hat{\mathbf{r}})|^2\propto\cos^2(\theta)$ as shown in Fig.~\ref{fig2}(b), which is exactly the classical well-known pattern of dipolar radiations~\cite{JACKSON_1998__Classical}. Relying on the same  logic, if we assume the $\cos^2(\theta)$-dependence angular pattern first, we would deduce that the dipolar radiations must be linearly polarized on the radiation plane.  That is, the reciprocity secures that the polarization and intensity distributions of dipolar radiations are interconnected rather than independent, and we can deduce either one from the other.  According to Eqs.~(\ref{coefficient-ED}) and (\ref{coefficient-ED-far}), the fully suppression of ED excitation and invisibility of the cylinder requires that $|\mathbf{E}_{\mathbb{I}}^1(-\hat{\mathbf{r}}_{1})|\cdot|\tilde{\mathbf{E}}_{\rm{ED}}(\hat{\mathbf{r}}_{1})|=|\mathbf{E}_{\mathbb{I}}^2(-\hat{\mathbf{r}}_{2})|\cdot|\tilde{\mathbf{E}}_{\rm{ED}}(\hat{\mathbf{r}}_{2})|$ and $\Delta\varphi=\pi$, which effectively leads to  $\mathbf{E}_z=0$.

It is worth mentioning that though Eqs.~(\ref{coefficient-ED}) and (\ref{coefficient-ED-far}) are equivalent, they are obtained from two different perspectives in the near and far field, respectively. Using the near-field technique, the calculations of the excitation coefficients of QNMs supported by other more sophisticated structures  inevitably rely on volume or surface integrations~\cite{LALANNE__LaserPhotonicsRev._Light,KRISTENSEN_2020_Adv.Opt.Photon.AOP_Modeling}, from which elegant formalism similar to  Eq.~(\ref{coefficient-ED}) and intuitive understandings are rarely obtainable; while employing our far-field model, the calculations are free from integrations and can always be obtained directly in the far field [according to Eq.~(\ref{coefficient-geometric})] despite any structural complexities, as long as the photonic structures are reciprocal. It enables us to manipulate the excitation coefficients and coherently control light-matter interactions, simply through tuning the properties of the incident plane waves. This is exactly the main point that distinguishes our framework for coherent controls from previous ones, as will be showcased in the following subsections.

\subsection*{Coherent excitations of individual QNMs}
We start with the simplest scenario of an individual QNM excited by two incident plane waves from different directions, for which according to  Eqs.~(\ref{expansion}) and (\ref{coefficient-geometric}) we have:
\begin{equation}
\label{expansion-indiviudal}
\begin{aligned}
\mathbf{E}_{\rm{sca}}(\hat{\mathbf{r}}) & \propto{\tilde{\mathbf{E}}}_{1}(\hat{\mathbf{{r}}})\sum_{n=1}^{2} |\mathbf{E}_{\mathbb{I}}^n(\hat{\mathbf{r}}_{\mathbb{I}}^{n})|\cdot|\tilde{\mathbf{E}}_1(-\hat{\mathbf{r}}_{\mathbb{I}}^{n})| \cos\left(\frac{1}{2}g_1^n\right)\exp({i\varphi_1^n}) \\
& =\alpha_1{\tilde{\mathbf{E}}}_{1}(\hat{\mathbf{{r}}}). 
\end{aligned}
\end{equation}
It tells that as long as $|\mathbf{E}_{\mathbb{I}}^1(\hat{\mathbf{r}}_{\mathbb{I}}^{1})|\cdot|\tilde{\mathbf{E}}_1(-\hat{\mathbf{r}}_{\mathbb{I}}^{1})| \cos\left(\frac{1}{2}g_1^1\right)=|\mathbf{E}_{\mathbb{I}}^2(\hat{\mathbf{r}}_{\mathbb{I}}^{2})|\cdot|\tilde{\mathbf{E}}_1(-\hat{\mathbf{r}}_{\mathbb{I}}^{2})| \cos\left(\frac{1}{2}g_1^2\right)$ and the phase contrast $\Delta\varphi_1^{1,2}=\varphi_1^1-\varphi_1^2=\pi$, the two excitation channels interfere fully destructively, with the QNM not been excited ($\alpha_1=0$) and thus the photonic structure being completely invisible. That is, for any incident wave, an extra wave from another direction can be shone to eliminate scatterings along all directions.  Similarly, when $\Delta\varphi_1^{1,2}=0$ the two channels would interfere constructively, maximizing the QNM excitation and thus also light-matter interactions.  Generally as explained, the amplitude (phase) of each channel can be controlled through the incident amplitudes and polarizations (phases) of all incident waves.

\begin{figure}[tp]
\centerline{\includegraphics[width=8.5cm]{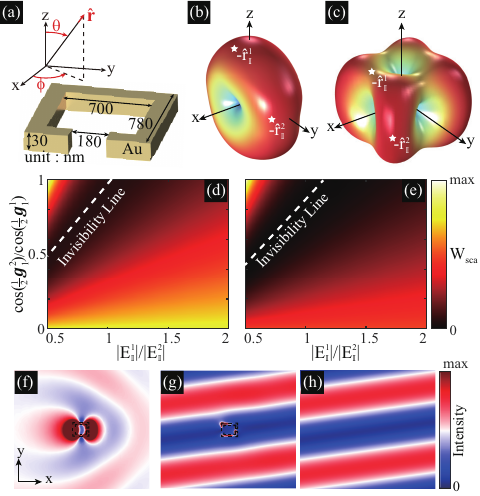}} \caption{\small (a) A gold SRR with all geometric parameters specified.  (b) \& (c) 3D far-field radiation patterns for the two low-order QNMs supported by the SRR, with complex eigenfrequencies $\tilde{\omega}_{1,2}$. For both scenarios two radiation directions (opposite to the incident directions) are pinpointed ($\star$). Dependence of $W_{\mathrm{sca}}$ on relative amplitude and polarization factors  $|\mathbf{E}_{\mathbb{I}}^1|/|\mathbf{E}_{\mathbb{I}}^2|$ and $\cos\left(\frac{1}{2}g_1^2\right)/\cos\left(\frac{1}{2}g_1^1\right)$ for the dipole-like QNM in (d) and for the 
quadrupole-like QNM in (e), where the dashed lines indicate parameters [predicted by Eq.~(\ref{expansion-indiviudal})] on which the particle becomes invisible. 
In (d): $\omega_{\mathbb{I}}=\mathrm{\mathbf{Re}}(\tilde{\omega}_{1})$;  $\hat{\mathbf{r}}_{\mathbb{I}}^{1}=(\theta,\phi)=(-\frac{17\pi}{20}, 0)$; $\hat{\mathbf{r}}_{\mathbb{I}}^{2}=(\frac{\pi}{2}, -\frac{11\pi}{20})$.  In (e): $\omega_{\mathbb{I}}=\mathrm{\mathbf{Re}}(\tilde{\omega}_{2})$;  $\hat{\mathbf{r}}_{\mathbb{I}}^{1}=(\theta,\phi)=(-\frac{3\pi}{4}, 0)$; $\hat{\mathbf{r}}_{\mathbb{I}}^{2}=(\frac{\pi}{2}, -\frac{3\pi}{4})$. Near-field distributions (in terms of total electric field intensity) on the \textbf{x}-\textbf{y} plane for three scenarios with $\omega_{\mathbb{I}}=\mathrm{\mathbf{Re}}(\tilde{\omega}_{1})$, respectively: (f) SRR scattering only incident wave $1$; (g) SRR scattering both incident waves; (h) two incident waves without SRR. In (f)-(h) incident and QNM radiation polarizations are fully matched [$\cos\left(\frac{1}{2}g_1^1\right)=\cos\left(\frac{1}{2}g_1^2\right)=1$] and in (g) \& (h) $|\mathbf{E}_{\mathbb{I}}^1|/|\mathbf{E}_{\mathbb{I}}^2|=1.064$.}\label{fig3}
\end{figure}

To confirm our QNM-based theoretical frame for coherent control of electromagnetic interactions, we begin with the widely employed seminal structure of split ring resonator (SRR) shown schematically in Fig.~\ref{fig3}(a). The SRR is gold-made (permittivity extracted from experimental data listed in Ref.~\cite{Johnson1972_PRB}) and numerical calculations are performed using COMSOL Multiphysics throughout this work. The far-field radiation patterns of the two low-order QNMs supported by the SRR are shown in Figs.~\ref{fig3}(b) and \ref{fig3}(c) with the directions opposite to the incident directions pinpointed ($\star$): the dipole-like QNM with complex eigenfrequency $\tilde{\omega}_{1}=(4.513 \times 10^{14}+1.394 \times 10^{14} \rm{i})$ rad/s, and the quadrupole-like QNM with $\tilde{\omega}_{2}=(1.005 \times 10^{15}+3.742 \times 10^{13} \rm{i})$ rad/s. The two QNMs are spectrally well-separated, and when the incident angular frequency $\omega_{\mathbb{I}}$ is close to a resonant frequency (real part of the QNM eigenfrequency), only the corresponding QNM would be excited. Coherent excitations for the two QNMs with two incident waves [incident at the corresponding resonant frequency $\omega_{\mathbb{I}}=\mathrm{\mathbf{Re}}(\tilde{\omega}_{1,2})$; incident directions in terms of polar and azimuthal angles $\hat{\mathbf{r}}_{\mathbb{I}}^{n}=(\theta,\phi)$ are specified in the corresponding figure captions, as is the case throughout the paper] are summarized respectively in Figs.~\ref{fig3}(d) and \ref{fig3}(e), and the coherent control is characterized by the evolutions of the angle-integrated scattered power $W_{\mathrm{sca}}$ (in the single-QNM regime where  only one QNM is effectively excited: $W_{\mathrm{sca}}\propto|\alpha_1|^2$). Since the full destructive interference ($W_{\mathrm{sca}}=|\alpha_1|^2=0$) is a characteristic feature of efficient coherent control, we have fixed $\Delta\varphi_1^{1,2}=\pi$ (this can be easily realized through tuning the incident phases) and show the dependence of $W_{\mathrm{sca}}$ on relative amplitude and polarization factors  $|\mathbf{E}_{\mathbb{I}}^1|/|\mathbf{E}_{\mathbb{I}}^2|$ and $\cos\left(\frac{1}{2}g_1^2\right)/\cos\left(\frac{1}{2}g_1^1\right)$. For both individual QNMs, as expected from Eq.~(\ref{expansion-indiviudal}), full destructive interferences are observed on straight invisibility lines $\cos\left(\frac{1}{2}g_1^2\right)/\cos\left(\frac{1}{2}g_1^1\right)  =|\tilde{\mathbf{E}}_1(-\hat{\mathbf{r}}_{\mathbb{I}}^{1})|/|\tilde{\mathbf{E}}_1(-\hat{\mathbf{r}}_{\mathbb{I}}^{2})| \cdot |\mathbf{E}_{\mathbb{I}}^1|/|\mathbf{E}_{\mathbb{I}}^2|$   [see dashed lines in Figs.~\ref{fig3}(d) and \ref{fig3}(e)]. 

To further visualize the effect of coherent control, we show in Figs.~\ref{fig3}(f)-\ref{fig3}(h) the near-field distributions (in terms of the total electric field intensity) on the \textbf{x}-\textbf{y} plane for three scenarios with respect to the dipole-like QNM excitations [$\omega_{\mathbb{I}}=\mathrm{\mathbf{Re}}(\tilde{\omega}_{1})$]: SRR with one incident wave; SRR with two incident waves; two incident waves without SRR (more details about incident properties can be found in the figure caption). It is clear that shining a second wave can fully eliminate the scattering [Figs.~\ref{fig3}(g)] that would be manifest with only one incident wave [Figs.~\ref{fig3}(f)], rendering the SRR effectively invisible [comparing Figs.~\ref{fig3}(g) and \ref{fig3}(h)]. Similar coherent control effects are also demonstrated for QNM absorption (see SI Appendix, Section 1), and for a more sophisticated scattering structure also in the single-QNM regime (see SI Appendix, Section 2).

\begin{figure}[tp]
\centerline{\includegraphics[width=9cm]{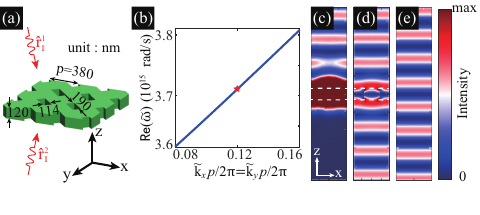}} \caption{(a) The square  PCS (periodicity $p=380$~nm; refractive index of the PCS is $2$; the background  refractive index is $1$) with all geometric parameters specified. (b)  One dispersion curve along the direction $\tilde{\mathbf{k}}_x=\tilde{\mathbf{k}}_y$ of the PCS with one Bloch QNM pinpointed ($\star$): $\tilde{\omega}_{3}=(3.71 \times 10^{15}+9.13 \times 10^{12} \rm{i})$ rad/s and $\tilde{\mathbf{k}}_xp/2\pi=\tilde{\mathbf{k}}_yp/2\pi=\tilde{\mathbf{k}}_{\star}p/2\pi=0.12$. Near-field distributions (in terms of electric field intensity) on the \textbf{x}-\textbf{z} plane for three scenarios: (c) the PCS interacting with only one incident wave; (d) PCS interacting with both incident waves; (e) two incident waves without the PCS. In (c)-(e) incident and QNM radiation polarizations are fully matched [$\cos\left(\frac{1}{2}g_1^1\right)=\cos\left(\frac{1}{2}g_1^2\right)=1$] and in (d) \& (e) $|\mathbf{E}_{\mathbb{I}}^1|/|\mathbf{E}_{\mathbb{I}}^2|=1$.}\label{fig4}
\end{figure}

Since our framework is built on the fundamental principle of reciprocity, it is applicable not only to finite scattering bodies, but also to infinitely extended structures.  We proceed to apply our model to infinitely-extended photonic crystal slabs (PCSs), which support Bloch QNMs characterized by complex eigenfrequencies $\tilde{\omega}$ and real in-plane wavevectors  $\tilde{\mathbf{k}}$ ~\cite{JOANNOPOULOS_2008__Photonic}. We study a  square PCS shown in  Fig.~\ref{fig4}(a) (periodicity $p=380$~nm; refractive index $2$; embedded in free background of refractive index $1$), with part of the band diagram ($\tilde{\mathbf{k}}_x=\tilde{\mathbf{k}}_y$) shown in Fig.~\ref{fig4}(b). One Bloch QNM is pinpointed ($\star$) in Fig.~\ref{fig4}(b), with $\tilde{\omega}_{3}=(3.71 \times 10^{15}+9.13 \times 10^{12} \rm{i})$ rad/s and $\tilde{\mathbf{k}}_xp/2\pi=\tilde{\mathbf{k}}_yp/2\pi=\tilde{\mathbf{k}}_{\star}p/2\pi=0.12$. There is one outstanding difference between QNMs supported by finite scattering bodies and infinitely extended periodic structures: for the former the QNM can radiate to all directions on the momentum sphere, while for the latter there is only a finite number of open radiation directions that correspond to  different diffraction orders~\cite{JOANNOPOULOS_2008__Photonic,CHEN_2019__Singularities}. That is, the opposite incident direction $-\hat{\mathbf{r}}_{\mathbb{I}}$ should overlap with one of the open radiation directions to effectively excite Bloch QNMs. To coherently control the excitation of the QNM pinpointed in Fig.~\ref{fig4}(b), we shine two incident waves from opposite sides of the PCS with $\omega_{\mathbb{I}}=\mathrm{\mathbf{Re}}(\tilde{\omega}_{3})$ and incident wavevector $\mathbf{k}_{\mathbb{I}}=(\mathbf{k}_{\mathbb{I},x},\mathbf{k}_{\mathbb{I},y},\mathbf{k}_{\mathbb{I},z})=(\tilde{\mathbf{k}}_{\star},\tilde{\mathbf{k}}_{\star},\pm\sqrt{|\mathbf{k}_0|^2-2\tilde{\mathbf{k}}_{\star}^2})$,
where $\mathbf{k}_0$ is the wavevector in free space.  The effect of coherent suppression is shown respectively through near-field distributions in Figs.~\ref{fig4}(c)-\ref{fig4}(e): PCS with one incident wave, with two incident waves, and two incident waves without the PCS (more details about incident properties can be found in the figure caption).  It is clear that, similar to the finite scatter shown in Fig.~\ref{fig3}, introducing an extra incident wave can render the PCS invisible.

\subsection*{Coherent excitations of Multiple QNMs}
As a next step, we proceed to the more sophisticated scenario of simultaneous two-QNM excitations. In contrast to the single-QNM case where introducing one extra incident wave guarantees the full coherent control, for two QNMs three incident waves are required for a comprehensive manipulation, as will be demonstrated below.   When the first incident wave excites both QNMs, and the incident polarization of the second (third) incident wave is orthogonal to the radiation polarization of the second (first) QNM opposite to the corresponding incident direction, that is  $\cos\left(\frac{1}{2}g_2^2\right)=0$ [$\cos\left(\frac{1}{2}g_1^3\right)=0$], according to Eqs.~(\ref{expansion}) and (\ref{coefficient-geometric}) we have (incident waves 2 and 3 do not excite QNM 2 and 1, respectively):
\begin{equation}
\label{expansion-two}
\begin{aligned}
&\mathbf{E}_{\rm{sca}}(\hat{\mathbf{r}}) \propto\\
& {\tilde{\mathbf{E}}}_{1}(\hat{\mathbf{{r}}})\sum_{n=1,2} |\mathbf{E}_{\mathbb{I}}^n(\hat{\mathbf{r}}_{\mathbb{I}}^{n})|\cdot|\tilde{\mathbf{E}}_1(-\hat{\mathbf{r}}_{\mathbb{I}}^{n})| \cos\left(\frac{1}{2}g_1^n\right)\exp({i\varphi_1^n})\\
 + & {\tilde{\mathbf{E}}}_{2}(\hat{\mathbf{{r}}})\sum_{n=1,3} |\mathbf{E}_{\mathbb{I}}^n(\hat{\mathbf{r}}_{\mathbb{I}}^{n})|\cdot|\tilde{\mathbf{E}}_2(-\hat{\mathbf{r}}_{\mathbb{I}}^{n})| \cos\left(\frac{1}{2}g_2^n\right)\exp({i\varphi_2^n})\\
= & \alpha_1{\tilde{\mathbf{E}}}_{1}(\hat{\mathbf{{r}}})+\alpha_2{\tilde{\mathbf{E}}}_{2}(\hat{\mathbf{{r}}}).
\end{aligned}
\end{equation}
Comparing Eqs.~(\ref{expansion-indiviudal}) and (\ref{expansion-two}), it is easy to notice that incident waves  $1$ and $2$ (incident waves $1$ and $3$) coherently excite QNM $1$ through manipulating $\alpha_1$ (QNM $2$ through $\alpha_2$) independently, and thus both QNM excitations can be selectively suppressed or enhanced. The only difference, compared to the single-QNM scenario, is that $\cos\left(\frac{1}{2}g_2^2\right)=0$  and $\cos\left(\frac{1}{2}g_1^3\right)=0$ lock respectively the polarizations of incident waves $2$ and $3$, but still  there is an extra degree of incident field amplitudes to tune to manipulate the excitation coefficients. It is worth mentioning that Eq.~(\ref{expansion-two}) would become more complicated and thus less elegant if we do not lock the polarizations of the incident waves  $2$ and $3$ [$\cos\left(\frac{1}{2}g_2^2\right)\neq0$  and $\cos\left(\frac{1}{2}g_1^3\right)\neq0$] or there are more than two QNMs simultaneously excited. The complexities originate from the increasing number of excitation channels and the inevitable involvement of geometric phase terms~\cite{WANG_2024_Phys.Rev.Lett._Geometric}, which would then turn the coherent control into more or less an engineering problem. In this study we have confined our discussions to the two-QNM excitation scenario, similar to the double-slit experiment from which all central principles can be revealed. 

\begin{figure}[tp]
\centerline{\includegraphics[width=9cm]{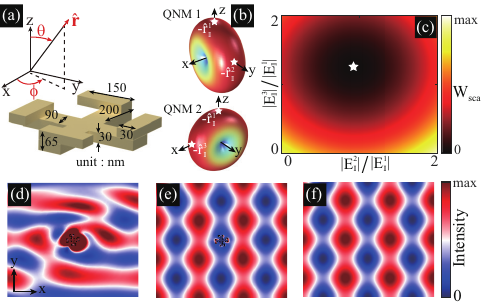}} \caption{(a) A gold structure that exhibits four-fold rotation symmetry and all its geometric parameters are specified. (b) 3D far-field radiation patterns for the two supported degenerate QNMs, with three directions opposite to the incident directions being pinpointed. Incident waves $2$ and $3$ do not contribute to the excitations of QNMs $2$ and $1$, respectively. (c) Dependence of $W_{\mathrm{sca}}$ on  $|\mathbf{E}_{\mathbb{I}}^2|/|\mathbf{E}_{\mathbb{I}}^1|$ and $|\mathbf{E}_{\mathbb{I}}^3|/|\mathbf{E}_{\mathbb{I}}^1|$. The Eq.~(\ref{expansion-two})-predicted point of invisiblity ($|\mathbf{E}_{\mathbb{I}}^2|/|\mathbf{E}_{\mathbb{I}}^1|=0.9$ and $|\mathbf{E}_{\mathbb{I}}^3|/|\mathbf{E}_{\mathbb{I}}^1|=1.3$) is marked ($\star$). The information of the three incident waves is: $\omega_{\mathbb{I}}=\mathrm{\mathbf{Re}}(\tilde{\omega}_{4})$;  $\hat{\mathbf{r}}_{\mathbb{I}}^{1}=(\pi, 0)$; $\hat{\mathbf{r}}_{\mathbb{I}}^{2}=(\pi/2,-\pi/2)$; $\hat{\mathbf{r}}_{\mathbb{I}}^{3}=(\pi/2,\pi)$. Near-field distributions (in terms of total electric field intensity) on the \textbf{x}-\textbf{y} plane for three scenarios, respectively: (d) The particle scattering only incident waves $1$ and $2$; (e) The particle scattering all three incident waves; (f) all three incident waves without the particle.}\label{fig5}
\end{figure}

To verify the above, we now turn to the gold particle shown in Fig.~\ref{fig5}(a), the rotation symmetry of which secures a pair of degenerate QNMs with shared eigenfrequency $\tilde{\omega}_{4}=(1.01 \times 10^{15}+9.61 \times 10^{13} \rm{i})$ rad/s. The far-field radiation patterns of both QNMs are shown in Fig.~\ref{fig5}(b), which are identical except for a $90^{\circ}$ rotation with respect to each other, as is required by the symmetry of the particle. The coherent excitation for the two degenerate QNMs with three incident [$\omega_{\mathbb{I}}=\mathrm{\mathbf{Re}}(\tilde{\omega}_{4})$] waves from different directions [pinpointed in Fig.~\ref{fig5}(b)] is summarized in Fig.~\ref{fig5}(c) through the dependence of $W_{\mathrm{sca}}$ on  $|\mathbf{E}_{\mathbb{I}}^2|/|\mathbf{E}_{\mathbb{I}}^1|$ and $|\mathbf{E}_{\mathbb{I}}^3|/|\mathbf{E}_{\mathbb{I}}^1|$, with fixed phase contrast $\Delta\varphi_1^{1,2}=\Delta\varphi_2^{1,3}=\pi$ (obtainable through tuning the phases of incident waves $2$ and $3$, respectively).  As expected from Eq.~(\ref{expansion-two}), simultaneous suppressions of both QNMs are obtainable with properly selected incident amplitudes, achieving invisibility at exactly the theoretically predicted ($\star$) position. The invisibility is further shown in Figs.~\ref{fig5}(d)-\ref{fig5}(f), respectively for two incident waves $1$ and $2$, for three incident waves with and without the scatterer. 
Similar coherent control effects are also demonstrated for another two-QNM configuration without mode degeneracy, with the QNMs being spectrally close and thus can be simultaneously excited (see SI Appendix, Section 3), and for infinitely extended PCS (see SI Appendix, Section 4).



\subsection*{Coherent controls of directional scattering intensity and polarization}
For coherent excitations of two QNMs, we rewrite Eq.~(\ref{expansion-two}) as: 
\begin{equation}
\label{expansion-directional}
\mathbf{E}_{\rm{sca}}(\hat{\mathbf{r}})\propto\alpha_1{\tilde{\mathbf{E}}}_{1}(\hat{\mathbf{{r}}})+\alpha_2{\tilde{\mathbf{E}}}_{2}(\hat{\mathbf{{r}}})={\tilde{\mathbf{E}}}_{a}(\hat{\mathbf{{r}}})+{\tilde{\mathbf{E}}}_{b}(\hat{\mathbf{{r}}}).
\end{equation}
For an arbitrary scattering direction $\hat{\mathbf{r}}$, we use points $\mathrm{R}_{1,2}$ on the Poincaré sphere to denote the radiation polarizations of the corresponding QNMs [$\mathrm{R}_{1,2}$ are exactly $\mathrm{Q}_{1,2}$  in Eq.~(\ref{coefficient-geometric}) when  $\hat{\mathbf{r}}$ is opposite to the incident direction].  When the QNM radiation polarizations along a direction are identical (overlapped $\mathrm{R}_{1,2}$), phases and amplitudes of $\alpha_{1,2}$ can be tuned accordingly [through tuning incident amplitudes and phases according to Eq.~(\ref{expansion-two})] to fully eliminate the directional scattering [${\tilde{\mathbf{E}}}_{a}(\hat{\mathbf{{r}}})=-{\tilde{\mathbf{E}}}_{b}(\hat{\mathbf{{r}}})$]. This essentially adds a novel QNM perspective for the dynamic research direction of generalized Kerker scattering, for which directional scattering  suppression (in particular along the backward direction) is conventionally interpreted from a perspective of multipolar parities and interferences~\cite{Kerker1983_JOSA,LIU_2018_Opt.Express_Generalized}. 

\begin{figure}[tp]
\centerline{\includegraphics[width=9cm]{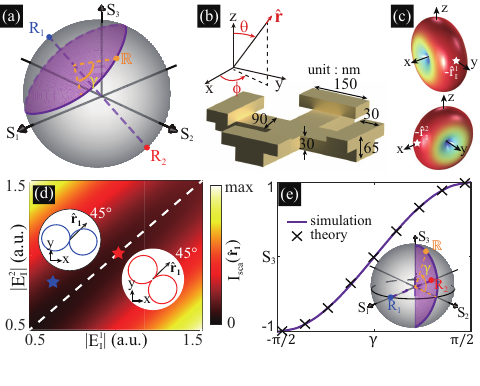}} \caption{\small (a) The final polarization $\mathbb{R}$ of the superimposed fields of orthogonal polarizations $\mathrm{R}_{1,2}$.  $\mathrm{R}$ locates on a circle perpendicular to diameter $\mathrm{R}_{1}\mathrm{R}_{2}$, and is parameterized by $\beta$ and $\gamma$. (b) A gold particle that supports a pair of degenerate QNMs, with all its geometric parameters specified. (c) 3D far-field radiation patterns for the two supported degenerate QNMs, with two directions opposite to the incident directions being pinpointed.
(d) The dependence of $I_{\rm{sca}}(\hat{\mathbf{r}}_1)$ on the incident amplitudes with fixed $\gamma=\pi$ [$I_{\rm{sca}}(\hat{\mathbf{r}}_1)=0$ on the dashed line]. Two points are pinpointed [locations are (1,1) and (0.6,0.8) for the red and blue stars, respectively], with the  corresponding angular scattering patterns on the \textbf{x}-\textbf{y} plane shown as insets. (e) Evolutions of the scattering polarizations (in terms of $S_3$) along +$\mathbf{z}$ for a fixed incident amplitude ratio ($\beta=1$) while varying phase contrast ($\gamma$). The inset of (e) shows the corresponding trajectory of the scattering polarizations on the Poincaré sphere.
For the incident waves: $\omega_{\mathbb{I}}=\mathrm{\mathbf{Re}}(\tilde{\omega}_{5})$;  $\hat{\mathbf{r}}_{\mathbb{I}}^{1}=(\pi/2,-\pi/2)$; $\hat{\mathbf{r}}_{\mathbb{I}}^{2}=(\pi/2,\pi)$.
}\label{fig6}
\end{figure}

If the radiation polarizations are distinct (non-overlapping $\mathrm{R}_{1,2}$) along $\hat{\mathbf{r}}$, though the directional scattering cannot be possibly eliminated, the phases and amplitudes of $\alpha_{1,2}$ can be tuned accordingly [based on  Eq.~(\ref{expansion-two})] to obtain arbitrary directional polarizations~\cite{RAMACHANDRAN_1961}.  To be specific, we define between the two contributing channels the relative amplitude as $\beta=|{\tilde{\mathbf{E}}}_{a}(\hat{\mathbf{{r}}})|/|{\tilde{\mathbf{E}}}_{b}(\hat{\mathbf{{r}}})|$ and relative phase between ${\tilde{\mathbf{E}}}_{a}(\hat{\mathbf{{r}}})$ and ${\tilde{\mathbf{E}}}_{b}(\hat{\mathbf{{r}}})$ as (adopting the Pancharatnam connection~\cite{RAMACHANDRAN_1961,Berry1987_JMP}) $\gamma=\mathrm{Arg}[{\tilde{\mathbf{E}}}_{a}(\hat{\mathbf{{r}}})\cdot {\tilde{\mathbf{E}}}_{b}^{*}(\hat{\mathbf{{r}}})]$ (* denote complex conjugation and $\mathrm{Arg}$ means the argument of complex numbers).  For the special scenario of orthogonal QNM radiation polarizations ($\mathrm{R}_{1,2}$ are diametrically opposite antipodal points on the Poincaré sphere), the final polarization state (denoted by point $\mathbb{R}$) for $\mathbf{E}_{\rm{sca}}(\hat{\mathbf{r}})$ is shown in Fig.~\ref{fig6}(a).  Except for the extreme scenario that only one QNM is excited ($\beta=0$ or $\infty$),  $\mathbb{R}$ locates on a circle on the Euclidian plane that is perpendicular to the diameter $\mathrm{R}_{1}\mathrm{R}_{2}$:  $\beta=\cos(\frac{1}{2}\overset{\smash{\raisebox{-0.1ex}{\tikz\draw (0,0) arc[start angle=110,end angle=70,radius=0.6cm];}}}{\mathbb{R} \mathrm{R}}_{1})/\cos(\frac{1}{2}\overset{\smash{\raisebox{-0.1ex}{\tikz\draw (0,0) arc[start angle=110,end angle=70,radius=0.6cm];}}}{\mathbb{R} \mathrm{R}}_{2})$; $\gamma$ is the in-plane azimuthal angle of $\mathbb{R}$ [see Fig.~\ref{fig6}(a)].  For the more general scenario of non-overlapping $\mathrm{R}_{1,2}$ that are not necessarily antipodal, refer to SI Appendix, Section 5.

To demonstrate coherent control of directional scattering intensity and polarizations,  we adopt the simplest configuration to shine two incident waves with fixed polarizations [$\cos\left(\frac{1}{2}g_1^2\right)=0$  and $\cos\left(\frac{1}{2}g_2^1\right)=0$], making sure that each incident wave would excite only one QNM (incident waves 1 and 2 would excite solely QNMs 1 and 2, respectively). Then according to Eq.~(\ref{expansion-two}), for Eq.~(\ref{expansion-directional}) we have explicitly:
\begin{equation}
\label{expansion-two-simplified}
\begin{aligned}
& \alpha_1=|\mathbf{E}_{\mathbb{I}}^1(\hat{\mathbf{r}}_{\mathbb{I}}^{1})|\cdot|\tilde{\mathbf{E}}_1(-\hat{\mathbf{r}}_{\mathbb{I}}^{1})| \cos\left(\frac{1}{2}g_1^1\right)\exp({i\varphi_1^1});\\
& \alpha_2=|\mathbf{E}_{\mathbb{I}}^2(\hat{\mathbf{r}}_{\mathbb{I}}^{2})|\cdot|\tilde{\mathbf{E}}_2(-\hat{\mathbf{r}}_{\mathbb{I}}^{2})| \cos\left(\frac{1}{2}g_2^2\right)\exp({i\varphi_2^2}),
\end{aligned}
\end{equation}
which tells that $\beta$ and $\gamma$ can be freely tuned through tailoring incident amplitudes and phases, respectively. For verifications, we then study another gold particle [shown in Fig.~\ref{fig6}(b)] that also supports a pair of degenerate QNMs [$\tilde{\omega}_{5}=(1.192 \times 10^{15}+1.307 \times 10^{14} \rm{i})$ rad/s]. The 3D scattering patterns of both QNMs are shown in  Fig.~\ref{fig6}(c), where the directions opposite to the incident directions are pinpointed.  Firstly we have identified a direction $\hat{\mathbf{r}}_1$ on the \textbf{x}-\textbf{y} plane with $\phi=\pi/4$ along which the QNM radiation polarizations are identical. In Fig.~\ref{fig6}(d) we show the dependence of angular scattering intensity along $\hat{\mathbf{r}}_1$ [$I_{\rm{sca}}(\hat{\mathbf{r}}_1)$] on the incident amplitudes with fixed $\gamma=\pi$.  As expected from Eqs.~(\ref{expansion-directional}) and (\ref{expansion-two-simplified}),  the points of $I_{\rm{sca}}(\hat{\mathbf{r}}_1)=0$ locate on a line [see the dashed line in Fig.~\ref{fig6}(d)]. We have further chosen two points (one on-line and one off-line) and show the corresponding angular scattering patterns on the \textbf{x}-\textbf{y} plane as insets of Fig.~\ref{fig6}(d), which further verifies that tuning properly the incident properties would result in directional scattering elimination. 

We have also identified another direction $\hat{\mathbf{r}}_2$ (the $+\mathbf{z}$ direction) along which the QNM radiation polarizations are orthogonal (both linearly polarized along $\mathbf{x}$ and $\mathbf{y}$ axes, respectively) and thus $\mathrm{R}_{1,2}$ are antipodal [see the inset of Fig.~\ref{fig6}(e)]. The evolutions of the scattering polarizations (in terms of $S_3$) along $\hat{\mathbf{r}}_2$ for a fixed incident amplitude ratio ($\beta=1$) while varying phase contrast ($\gamma$) are showcased in Fig.~\ref{fig6}(e), which agree perfectly with the predictions of Eqs.~(\ref{expansion-directional}) and (\ref{expansion-two-simplified}), and the general picture shown in Fig.~\ref{fig6}(a): $S_3=\sin(\gamma)$. The directional scattering can be circularly-polarized ($S_3=\pm1$ with $\gamma=\pm\pi/2$), linearly-polarized ($S_3=0$ with $\gamma=0$), and elliptically-polarized for other phase contrast [see the inset of Fig.~\ref{fig6}(e)]

\section*{Conclusions and outlook}

In conclusion, we have merged two sweeping concepts of coherent controls and QNMs to establish a framework for efficient manipulations of light-matter interactions. It is revealed that for both finite scatterers and infinite periodic structures, introducing extra beams incident from different directions can coherently control the excitations of the QNMs supported, thus enabling  flexible manipulations of both total and directional scatterings. In our framework, all properties of incident waves (amplitudes, phases and polarizations) can be thoroughly exploited in a synchronous manner to architect the excitation coefficients of targeted QNMs with high precision, thus opening new avenues for the vibrant discipline of electromagnetic coherent controls. 

In this study, we have confined the incident waves to be plane waves. For more sophisticated structured incident waves that can be expanded into plane waves, our framework can be further generalized to be more widely applicable through replacing the discrete summation over incident wave index $m$ in Eq.~(\ref{expansion}) by a continuous integration over incident momentum. An obvious limitation of our model is that it is applicable only to reciprocal structures, and with broken reciprocity (by magnetism, nonlinearities and/or time modulations) currently it is not yet known whether or not an elegant far-field geometric  formalism similar to Eq.~(\ref{coefficient-geometric}) exists. Considering the ubiquity of coherent controls and QNMs in almost every branch of wave physics, the framework we establish can provide a fertile platform for investigations into not only light-matter interactions, but many other sorts of interactions involving waves of various forms, spanning subjects beyond optics and photonics to acoustics and mechanics, plasma physics and even gravitational wave physics.

\section*{Materials and Methods}

Throughout this work, for numerical calculations of both the source-free QNMs (properties including complex eigenfrequencies, near-field distributions and far-field radiation and polarization distributions) and scattering properties with different incident sources, we use the commercial software COMSOL Multiphysics. Analytical results are obtained directly from the equations listed in the manuscript. For gold structures,  the experimental permittivity is fitted by the Drude model $\varepsilon(\omega_{\mathbb{I}})=1-\omega_p^2 / \omega_{\mathbb{I}}\left(\omega_{\mathbb{I}}+i \omega_c\right)$, where $\omega_p \approx 1.37 \times 10^{16} \mathrm{rad}/\mathrm{s}$ is the plasma frequency and $\omega_c \approx 8.17 \times 10^{13} \mathrm{rad}/\mathrm{s}$ is the collision frequency. To avoid interrupting the logic flow in the main text, detailed geometric and optical parameters (incident directions and polarizations \textit{etc}.) are specified in either the figure directly or in figure captions. 

\emph{Acknowledgment}:  This research was funded by the National
Natural Science Foundation of China (Grants No. 12274462,
No.11674396, and No. 11874426) and several other projects of
Hunan Province (Projects No. 2024JJ2056 and No. 2023JJ10051). W. L. is grateful to M. V. Berry and J. F. Zhang for invaluable correspondences.


\bibliography{References_scattering8}




\onecolumngrid
\clearpage


{\centering
  \noindent\textbf{\large{Supplementary Information for:}}
\\\vspace{0.1cm}
\noindent\textbf{\large{``Quasi-normal modes empowered coherent control of electromagnetic interactions"}}
\\\bigskip


Jingwei Wang$^{1}$, Pengxiang Wang$^{1}$, Chaofan Zhang$^{2,3,*}$, Yuntian Chen$^{1,4,\dag}$, and Wei Liu$^{2,3,\ddagger}$
\\

\small{$^1$ \emph{School of Optical and Electronic Information, Huazhong University of Science and Technology, Wuhan, Hubei 430074, P. R. China. }}\\
\small{$^2$ \emph{College for Advanced Interdisciplinary Studies, National University of Defense Technology, Changsha 410073, P. R. China.}}\\
\small{$^3$ \emph{Nanhu Laser Laboratory and Hunan Provincial Key Laboratory of Novel Nano-Optoelectronic Information Materials and Devices,
National University of Defense Technology, Changsha 410073, P. R. China.}}\\
\small{$^4$ \emph{Wuhan National Laboratory for Optoelectronics, Huazhong University of Science and Technology, Wuhan, Hubei 430074, P. R. China.}}\\
$^{*}$~c.zhang@nudt.edu.cn; $^{\dag}$~~yuntian@hust.edu.cn; $^{\ddagger}$~wei.liu.pku@gmail.com\\ \vspace{0.3cm}

{\leftskip=0pt \rightskip=0pt plus 0cm

This supporting information includes the following five sections: ((\textbf{\uppercase\expandafter{\romannumeral1}}). Coherent control of the QNM absorption; (\textbf{\uppercase\expandafter{\romannumeral2}}). Coherent excitations of individual QNMs supported by a pair of coupled SRRs; (\textbf{\uppercase\expandafter{\romannumeral3}}). Coherent excitations of two non-degenerate QNMs supported by a SRR; (\textbf{\uppercase\expandafter{\romannumeral4}}). Coherent excitations of two degenerate QNMs supported by a PCS;  (\textbf{\uppercase\expandafter{\romannumeral5}}). The polarization of a superimposed wave from two non-orthogonally polarized waves.\\\bigskip 

\setcounter{equation}{0}
\setcounter{figure}{0}
\newcounter{sfigure}
\setcounter{sfigure}{1}
\setcounter{table}{0}
\renewcommand{\theequation}{S\arabic{equation}}

 \renewcommand\thefigure{S{\arabic{figure}}}
\renewcommand{\thesection}{S\arabic{section}}

\twocolumngrid

\section{(\textbf{\uppercase\expandafter{\romannumeral1}}). Coherent control of the QNM absorption.}
\label{berry-dennis model}

In the main article, we demonstrate the effects of coherent control through the total scattered powers, and here we showcase the effects through the total absorbed power ($W_{\mathrm{abs}}$). The structures shown in Figs.~\ref{figs-1}(a) and \ref{figs-1}(b) are identical to those in Fig. 3(a) and Fig. 5(a), respectively.  The results shown in Figs.~\ref{figs-1}(c) and \ref{figs-1}(d) are the same respectively as those in Fig. 3(d) and Fig. 5(c), except now that it is $W_{\mathrm{abs}}$ rather than $W_{\mathrm{sca}}$. It is clear that the absorption can be fully suppressed with tailored incident properties (on the dashed line and the marked star).

\begin{figure}
\centerline{\includegraphics[width=9cm]{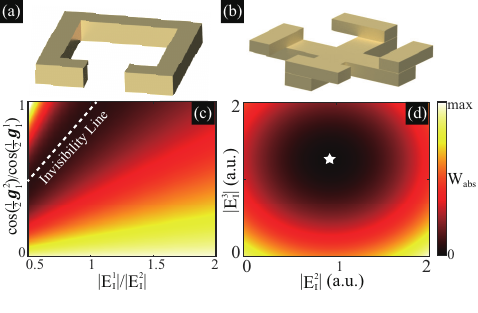}} \caption{\small (a) and (b) are identical to those in Fig. 2(a) and Fig. 5(a). (c) and (d) are the same respectively as those in Fig. 2(d) and Fig. 5(c) of the main article, except now that it is the dependence of $W_{\mathrm{abs}}$ rather than $W_{\mathrm{sca}}$. In (c) and (d) the absorption elimination positions are marked.}
\label{figs-1}
\end{figure}

\section{(\textbf{\uppercase\expandafter{\romannumeral2}}).  Coherent excitations of individual QNMs supported by a pair of coupled SRRs.}

\begin{figure}
\centerline{\includegraphics[width=9cm]{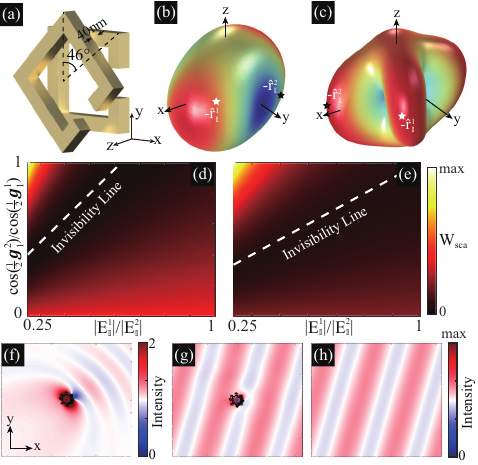}} \caption{\small (a) A gold SRR pair with geometric parameters specified [each SRR is the same as that shown in Fig. 3(a)].  (b) \& (c) 3D far-field radiation patterns for the two lowest-order QNMs supported by the SRR, with complex eigenfrenquencies $\tilde{\omega}_{6,7}$. For both scenarios, two radiation directions (opposite to the incident directions) are pinpointed ($\star$). Dependence of $W_{\mathrm{sca}}$ on relative amplitude and polarization factors  $|\mathbf{E}_{\mathbb{I}}^1|/|\mathbf{E}_{\mathbb{I}}^2|$ and $\cos\left(\frac{1}{2}g_1^2\right)/\cos\left(\frac{1}{2}g_1^1\right)$ for the dipole-like QNM in (d) and for the 
quadrupole-like QNM in (e), where the dashed lines indicate parameters [predicted by Eq.~(5)] on which the particle becomes invisbile. 
In (d): $\omega_{\mathbb{I}}=\mathrm{\mathbf{Re}}(\tilde{\omega}_{6})$;  $\hat{\mathbf{r}}_{\mathbb{I}}^{1}=(\theta,\phi)=(\frac{11\pi}{20}, -\frac{7\pi}{10})$; $\hat{\mathbf{r}}_{\mathbb{I}}^{2}=(\frac{\pi}{2}, -\frac{3\pi}{20})$.  In (e): $\omega_{\mathbb{I}}=\mathrm{\mathbf{Re}}(\tilde{\omega}_{7})$;  $\hat{\mathbf{r}}_{\mathbb{I}}^{1}=(\theta,\phi)=(\frac{\pi}{2}, -\frac{2\pi}{3})$; $\hat{\mathbf{r}}_{\mathbb{I}}^{2}=(\frac{\pi}{2}, -\frac{\pi}{20})$. Near-field distributions (in terms of total electric field intensity) on the \textbf{x}-\textbf{y} plane for three scenarios with $\omega_{\mathbb{I}}=\mathrm{\mathbf{Re}}(\tilde{\omega}_{6})$, respectively: (f) SRR pair scattering only incident wave $1$; (g) SRR pair scattering both incident waves; (h) two incident waves without SRR pair. In (f)-(h) incident and QNM radiation polarizations are fully matched [$\cos\left(\frac{1}{2}g_1^1\right)=\cos\left(\frac{1}{2}g_1^2\right)=1$] and in (g) \& (h) $|\mathbf{E}_{\mathbb{I}}^1|/|\mathbf{E}_{\mathbb{I}}^2|=0.667$.}
\label{figs-2}
\end{figure}

We also show coherent control for QNMs supported by a pair of coupled split ring resonators shown schematically in Fig.~\ref{figs-2}(a). Each SRR is the same as that shown in Fig. 3(a). The far-field radiation patterns of the two lowest-order QNMs supported by the SRR are shown in Figs.~\ref{figs-2}(b) and \ref{figs-2}(c) with the directions opposite to the incident directions pinpointed ($\star$): the dipole-like QNM with complex eigenfrequency $\tilde{\omega}_{6}=(2.377 \times 10^{14}+1.475 \times 10^{13} \rm{i})$ rad/s, and the quadrupole-like QNM with $\tilde{\omega}_{7}=(1.34 \times 10^{15}+6.786 \times 10^{13} \rm{i})$ rad/s. The two QNMs are spectrally well-separated, and when the incident angular frequency $\omega_{\mathbb{I}}$ is close to a resonant frequency (real part of the QNM eigenfrequency), only the corresponding QNM would be excited. Coherent excitations for the two QNMs with two incident waves [incident at the corresponding resonant frequency $\omega_{\mathbb{I}}=\mathrm{\mathbf{Re}}(\tilde{\omega}_{6,7})$; incident directions are specified in the corresponding figure captions] are summarized respectively in Figs.~\ref{figs-2}(d) and \ref{figs-2}(e), and the coherent control is characterized by the evolutions of $W_{\mathrm{sca}}$ (in the single-QNM regime where  only one QNM is effectively excited: $W_{\mathrm{sca}}\propto|\alpha_1|^2$). Similar to the results shown in Figs. 3(d) and 3(e), we have fixed $\Delta\varphi_1^{1,2}=\pi$ show the dependence of $W_{\mathrm{sca}}$ on relative amplitude and polarization factors  $|\mathbf{E}_{\mathbb{I}}^1|/|\mathbf{E}_{\mathbb{I}}^2|$ and $\cos\left(\frac{1}{2}g_1^2\right)/\cos\left(\frac{1}{2}g_1^1\right)$. For both individual QNMs, full destructive interferences are observed on straight invisibility lines  [see dashed lines in Figs.~\ref{figs-2}(d) and \ref{figs-2}(e)]. 

To further visualize the effect of coherent control, we show in Figs.~\ref{figs-2}(f)-\ref{figs-2}(h) the near-field distributions (in terms of the total electric field intensity) on the \textbf{x}-\textbf{y} plane for three scenarios with respect to the dipole-like QNM excitations [$\omega_{\mathbb{I}}=\mathrm{\mathbf{Re}}(\tilde{\omega}_{6})$]: SRR pair with one incident wave; SRR pair with two incident waves; two incident waves without SRR pair (more details about incident properties can be found in the figure caption). It is clear that shining a second wave can fully eliminate the scattering [Figs.~\ref{figs-2}(g)] that would be manifest with only one incident wave [Figs.~\ref{figs-2}(f)], rendering the SRR pair effectively invisible [comparing  Figs.~\ref{figs-2}(g) and \ref{figs-2}(h)].

\begin{figure}
\centerline{\includegraphics[width=9cm]{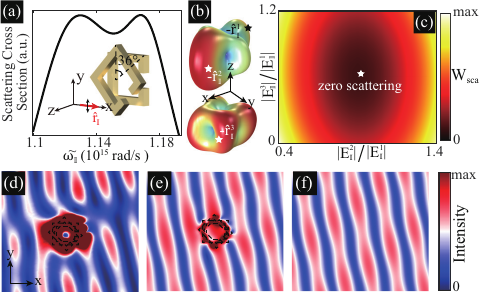}} \caption{\small (a) The scattering cross section spectra for the structure shown as inset, where two peaks are manifest that correspond to the two QNMs excited. The incident plane wave is propagating along +\textbf{x} direction and linearly polarized along \textbf{y} axis.  (b) 3D far-field radiation patterns for the two supported non-degenerate QNMs, with three directions opposite to the incident directions being pinpointed. (c) Dependence of $W_{\mathrm{sca}}$ on  $|\mathbf{E}_{\mathbb{I}}^2|/|\mathbf{E}_{\mathbb{I}}^1|$ and $|\mathbf{E}_{\mathbb{I}}^3|/|\mathbf{E}_{\mathbb{I}}^1|$. The Eq.~(6)-predicted point of invisibility ($|\mathbf{E}_{\mathbb{I}}^2|/|\mathbf{E}_{\mathbb{I}}^1|=0.875$ and $|\mathbf{E}_{\mathbb{I}}^3|/|\mathbf{E}_{\mathbb{I}}^1|=0.66$) is marked ($\star$). The information of the three incident waves is: $\hat{\mathbf{r}}_{\mathbb{I}}^{1}=(\pi/2, 0)$; $\hat{\mathbf{r}}_{\mathbb{I}}^{2}=(\pi/2,-4\pi/5)$; $\hat{\mathbf{r}}_{\mathbb{I}}^{3}=(\pi/2,-5\pi/8)$. Near-field distributions (in terms of total electric field intensity) on the \textbf{x}-\textbf{y} plane for three scenarios, respectively: (d) The particle scattering only incident waves $1$ and $2$; (e) The particle scattering all three incident waves; (f) all three incident waves without the particle.}
\label{figs-3}
\end{figure}

\section{(\textbf{\uppercase\expandafter{\romannumeral3}}). Coherent excitations of two non-degenerate QNMs supported by a SRR.}

We proceed to show coherent controls for a pair of non-degenerate QNMs supported by the SRR-pair [each SRR is identical to the one in Fig. 3(a)] shown in the inset of Fig.~\ref{figs-3}(a): for its scattering cross section spectra (incident plane wave is propagating along +\textbf{x} direction and linearly polarized along \textbf{y} axis) two peaks that correspond to the two QNMs are observed. The complex eigenfrequencies are $\tilde{\omega}_{8}=(1.115 \times 10^{14}+7.613 \times 10^{13} \rm{i})$ rad/s, and $\tilde{\omega}_{9}=(1.166 \times 10^{15}+5.945 \times 10^{13} \rm{i})$ rad/s. The two QNMs are spectrally close and can be simultaneously excited  when the incident frequency is close to either resonant frequency. The far-field radiation patterns of both QNMs are shown in Fig.~\ref{figs-3}(b), with the directions opposite to the incident directions pinpointed ($\star$). For incident frequency $\omega_{\mathbb{I}}=[\mathrm{\mathbf{Re}}(\tilde{\omega}_{8})+\mathrm{\mathbf{Re}}(\tilde{\omega}_{9})]/2$, the coherent excitation for the two non-degenerate QNMs with three incident waves from different directions is summarized in Fig.~\ref{figs-3}(c) through the dependence of $W_{\mathrm{sca}}$ on  $|\mathbf{E}_{\mathbb{I}}^2|/|\mathbf{E}_{\mathbb{I}}^1|$ and $|\mathbf{E}_{\mathbb{I}}^3|/|\mathbf{E}_{\mathbb{I}}^1|$, with fixed phase contrast $\Delta\varphi_1^{1,2}=\Delta\varphi_1^{1,3}=\pi$ (obtainable through tuning the phases of incident waves $2$ and $3$, respectively).  As expected from Eq.~(6), simultaneous suppressions of both QNMs are obtainable with proper selective incident amplitudes, achieving invisibility at exactly the theoretically predicted ($\star$) position. The invisibility is further shown in Figs.~\ref{figs-3}(d)-\ref{figs-3}(f), respectively for two incident wave, for three incident waves with and without the SRR-pair.

\section{(\textbf{\uppercase\expandafter{\romannumeral4}}). Coherent excitations of two degenerate QNMs supported by a PCS}

\begin{figure}
\centerline{\includegraphics[width=10cm]{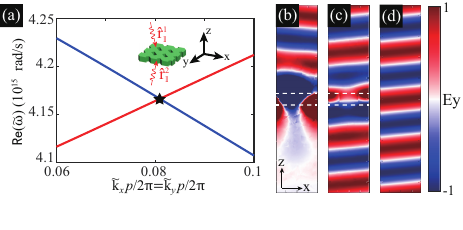}} \caption{\small (a)  The dispersion curves of the same PCS studied in Fig. 4(a) of the main article (see also the inset), now in another spectral regime, with one pair of degenerate Bloch QNMs marked by a $\star$: $\tilde{\omega}_{10}=(4.166 \times 10^{15}+7.774 \times 10^{12} \rm{i} )$ rad/s, $\tilde{\omega}_{11}=(4.166 \times 10^{15}+4.699 \times 10^{12} \rm{i} )$ rad/s and $\tilde{\mathbf{k}}_xp/2\pi=\tilde{\mathbf{k}}_yp/2\pi=\tilde{\mathbf{k}}_{\star}p/2\pi=0.081$. Near-field distributions on the x-y plane for three scenarios, respectively: (b) the PCS interacting with only incident wave $1$ that excits both Bloch QNMs; (c) PCS interacting with both incident waves with both QNMs unexcited; (d) both incident waves without the PCS. For the two incident waves: $\omega_{\mathbb{I}}=\mathrm{\mathbf{Re}}(\tilde{\omega}_{10})$; $\mathbf{k}_{\mathbb{I}}=(\mathbf{k}_{\mathbb{I},x},\mathbf{k}_{\mathbb{I},y},\mathbf{k}_{\mathbb{I},z})=(\tilde{\mathbf{k}}_{\star},\tilde{\mathbf{k}}_{\star},\pm\sqrt{|\mathbf{k}_0|^2-2\tilde{\mathbf{k}}_{\star}^2})$;
$\mathbf{k}_0$ is the wavevector in free space. In (c) \& (d) $|\mathbf{E}_{\mathbb{I}}^1|/|\mathbf{E}_{\mathbb{I}}^2|=1$.}\label{figs-4}
\end{figure}

For the same square PCS studied in Fig.4(a) of the main article [see also the inset of Fig.~\ref{figs-4}(a)], we have identified a pair of degenerate Bloch QNMs [marked in the band diagram shown in Fig.~\ref{figs-4}(a)] with eigenfrequencies $\tilde{\omega}_{10}=(4.166 \times 10^{15}+7.774 \times 10^{12} \rm{i} )$ rad/s and $\tilde{\omega}_{11}=(4.166 \times 10^{15}+4.699 \times 10^{12} \rm{i} )$ rad/s. Simultaneous suppressions of both QNMs and thus invisibility of the PCS are further shown in  Figs.~\ref{figs-4}(b)-\ref{figs-4}(d), respectively for one incident wave, and for both incident waves with and without the PCS.

\section{(\textbf{\uppercase\expandafter{\romannumeral5}}).  The polarization of a superimposed wave from two non-orthogonally polarized waves.}

For the more general scenario of non-overlapping $\mathrm{R}_{1,2}$ that are not necessarily antipodal, the representation of the final polarization on the Poincaré sphere ~\cite{RAMACHANDRAN_1961} is shown in Fig.~\ref{figs-5}: the great circle connecting $\mathrm{R}_{1}$ and $\mathrm{R}_{2}$ is colored red; the other great circle perpendicular to the red great circle and bisecting $\overset{\smash{\raisebox{-0.1ex}{\tikz\draw (0,0) arc[start angle=110,end angle=70,radius=0.9cm];}}}{\mathrm{R}_1\mathrm{R}_2}$ is colored blue; for fixed $\beta=\beta_0$, $\mathbb{R}$ locates on the yellow spherical circle with its center $\mathrm{\mathbf{A}}$ on the red great circle; for fixed $\gamma=\gamma_0$, $\mathbb{R}$ locates on the purple spherical arc with its center $\mathrm{\mathbf{B}}$ on the blue great circle.

\begin{figure}
\centerline{\includegraphics[width=5cm]{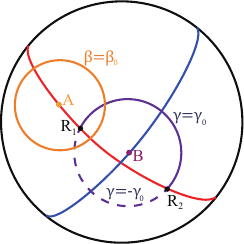}} \caption{\small Representations on the Poincaré sphere for the final polarization state $\mathbb{R}$  of  a superimposed wave from two non-orthogonally polarized waves of polarizations $\mathrm{R}_{1}$ and $\mathrm{R}_{2}$.}
\label{figs-5}
\end{figure}
%

\end{document}